\documentclass[aps,prl,showpacs,twocolumn,superscriptaddress]{revtex4}

\usepackage{amssymb}
\usepackage{graphicx}
\usepackage{dcolumn}
\usepackage{bm}
\usepackage{color}

\begin{document}

\title{Damping of Exciton Rabi Rotations by Acoustic Phonons in Optically Excited InGaAs/GaAs
Quantum Dots}

\author{A.~J.~Ramsay}
\email{a.j.ramsay@shef.ac.uk}
\affiliation{Department of Physics
and Astronomy, University of Sheffield, Sheffield, S3 7RH, United
Kingdom}

\author{Achanta~Venu~Gopal}
\affiliation{DCMP \& MS, Tata Institute of Fundamental Research, Mumbai 400 005, India}

\author{E.~M.~Gauger}
\affiliation{Department of Materials, University of Oxford, Oxford OX1 3PH, United Kingdom}

\author{A.~Nazir}
\affiliation{Department of Physics and Astronomy, University College London, London, WC1E 6BT, United Kingdom}

\author{B.~W.~Lovett}
\affiliation{Department of Materials, University of Oxford, Oxford OX1 3PH, United Kingdom}

\author{A.~M.~Fox}
\affiliation{Department of Physics and Astronomy, University of
Sheffield, Sheffield, S3 7RH, United Kingdom}

\author{M.~S.~Skolnick}
\affiliation{Department of Physics and Astronomy, University of
Sheffield, Sheffield, S3 7RH, United Kingdom}

\date{\today}

\begin{abstract}
We report experimental evidence identifying acoustic phonons as the principal source of the excitation-induced-dephasing (EID) responsible for the intensity damping of quantum dot excitonic Rabi rotations. The rate of EID is extracted from temperature dependent Rabi rotation measurements of the ground-state excitonic transition, and is found to be in close quantitative agreement with an acoustic-phonon model.
\end{abstract}
\maketitle

Semiconductor quantum dots provide nanoscale electronic confinement resulting in discrete energy levels, and an atom-like light-matter interaction.
Unlike natural atoms, quantum dots interact with their solid-state environment, resulting in a light-atom-continuum system with nonlinear dephasing dynamics. The effect of these interactions can be observed in excitonic Rabi rotations, where a picosecond control laser
drives a damped oscillation in the population inversion with increasing pulse-area \cite{Zrenner_nat,Stufler_prb2,RabiR,Wang_prb,Patton_prl}.

The cause of the intensity damping of excitonic Rabi rotations is the subject of intense debate. Three possible mechanisms have emerged: excitation of multi-exciton transitions, wetting layer and acoustic phonon mediated dephasing. Multi-exciton transitions damp the Rabi rotation when they are also excited by a  spectrally broad laser \cite{Patton_prl}. The wetting layer is a quantum well that occurs naturally in the growth of
quantum dots by the Stranski-Krastanov method, which have the high optical quality required to function as solid-state qubits. The wetting layer model \cite{VillasBoas_prl} attributes the damping to excitation-induced-dephasing (EID) caused by non-resonant excitation of hybridized states, where one carrier is confined to the dot and the other to the wetting layer \cite{Vasanelli_prl}, and is supported by experimental studies of Rabi rotations of
p-shell (excited state) excitons \cite{Wang_prb}. 

In the phonon model \cite{Forstner_prl,Vagov_prl,Gauger_prb}, the control laser modulates the electronic state of the dot at the Rabi frequency $\Omega$, and couples to  lattice vibrations via the deformation potential. A resonant exchange of energy occurs between the electronic polarization of the dot and the longitudinal acoustic (LA) phonon modes,
at a rate proportional to  the frequency response $K(\Omega)$
of the exciton-phonon bath coupling, dissipating the excitonic phase information
to the large phase-space of the LA-phonon modes. Considerable theoretical work has been reported on phonon mediated dephasing in the Rabi regime \cite{Forstner_prl,Vagov_prl,Gauger_prb,Axt_prb}, but there is as yet no experimental support for this work. 
In contrast to previous work on phonon dephasing \cite{Borri_prl}, this report focuses on the EID during coherent exciton manipulation, an optical-field dependent rate of dephasing, dominated by the phonon modes resonant with the Rabi frequency.

The importance of understanding EID is emphasized by recent demonstrations of full \cite{Press_nat}, and partial \cite{Berezovsky_sci,Ramsay_prl} picosecond control of single spins using charged exciton optical transitions. Optical spin control offers the exciting possibility of a semiconductor qubit that combines picosecond gate-times with microsecond coherence times. However, in practice, the gate fidelity will be limited by dephasing of the excitonic states occupied during the coherent control. Therefore, strategies for minimizing EID of such states will be key to achieving the $\sim10^3$ coherent operations that would be required for fault-tolerant spin quantum computation.

In this letter, we report compelling experimental evidence identifying acoustic phonons as the dominant source of intensity damping of excitonic Rabi rotations  in semiconductor quantum dots. We study the s-shell neutral exciton transition of single InGaAs/GaAs
self-assembled quantum dots using photocurrent detection. Rabi rotations are measured as a
function of temperature, and a characteristic parameter with units of time, termed the EID-time $K_2$, is extracted. This parameter expresses the rate of excitation-induced-dephasing, and is defined by $\Gamma_2=K_2\Omega^2$, where $\Gamma_2$ is the rate of pure dephasing and $\Omega$ the Rabi frequency. We find $K_2$ to be linear in temperature, with a gradient in quantitative agreement with a calculation based on an exciton-phonon model in the Born-Markov approximation. This provides strong evidence that acoustic phonons are responsible for EID of the fundamental excitonic transitions of  InGaAs/GaAs quantum dots.

The samples studied here are single InGaAs/GaAs dots embedded in
the intrinsic GaAs region of an n-i-Schottky diode emitting at 952 nm. Typically the dots have a base diameter of 20 nm, and a height of 3-4 nm.
The peak of the wetting layer emission is 861 nm.
A mode-locked Ti:sapphire laser provides a source of
150-fs pulses with a center wavelength of 951 nm at a repetition rate of 76 MHz. The
beam is passed through a pulse-shaper consisting of a 4F zero-dispersion compensator with an
adjustable slit in the masking plane.
To ensure that only the neutral exciton transition is excited, the pulses have a spectral width that is small compared with the 1.9-meV biexciton binding energy. The pulse is circularly polarized to further suppress two-photon absorption on the biexciton transition
\cite{Stufler_prb}. The sample sits in a cold-finger cryostat connected to
a measurement circuit for photocurrent detection \cite{Zrenner_nat}. A background photocurrent linear in
the power is subtracted from all of the data. The background signal is attributed to weak excitation of
other dots in the same mesa by light scattered within the sample \cite{Stufler_prb2}. Further details of the sample and setup can be found in
Ref. \cite{Boyle_prb}.

\begin{figure}
\begin{center}
\vspace{0.2 cm}
\includegraphics[scale=2]{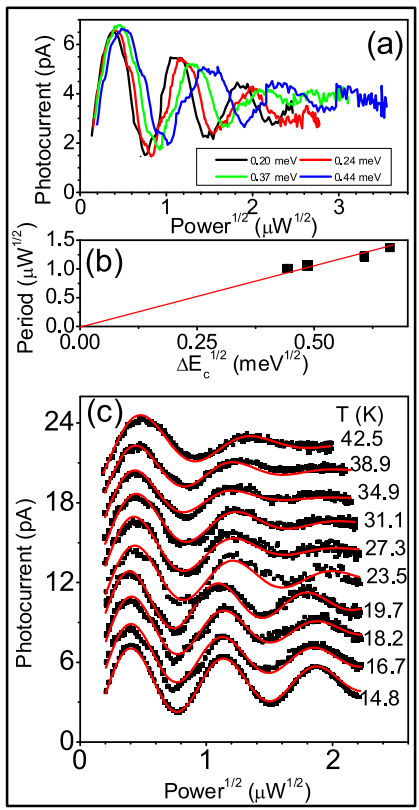}
\vspace{0.2 cm}
\end{center}
\caption{ Photocurrent versus square-root of
incident power for on-resonance excitation.
(a) Dependence of Rabi rotations on the spectral FWHM $\Delta E_c$ of the driving pulse.
(b) The period of the Rabi rotations shown in (a) versus $\sqrt{\Delta E_c}$.
(c) Temperature dependence of Rabi rotations, at a gate voltage of 0.6 V. The damping
of the Rabi rotation increases with temperature. The red-lines are fits to theory (see main text).
}\label{fig:RR}
\end{figure}

In Fig.~\ref{fig:RR}, we show two sets of Rabi rotation measurements. Here, the laser pulse is tuned on resonance with the neutral exciton transition (driving a coherent rotation between the
crystal-ground and neutral exciton states), and the final exciton population is recorded as an oscillation in
the photocurrent with increasing pulse-area ($\Theta=\int\Omega(t)dt$) of the optical field, which is
proportional to the square-root of the incident power. 
The amplitude of the oscillations decay with increasing excitation power.
It is this damping that limits the fidelity of qubit rotations using excitonic transitions,
and is the focus of this letter.

In the first set of data
a series of Rabi rotations for excitation pulses of various spectral widths was measured,
 and is presented in Fig.~\ref{fig:RR}(a). The device is at a temperature of 15~K, and held at a reverse bias of 0.6~V, where electron tunneling is slow ($\Gamma_{\rm e}^{\rm -1}>60~\mathrm{ps}$).
 This set of data has two main features. Firstly, the period of the oscillation is proportional to the square-root of the spectral full-width half-maximum of the pulse $\Delta E_c$,
 as plotted in Fig.~\ref{fig:RR}(b). This variation can be explained using a two-level atom model \cite{Allen_book}, where the angle of the
Rabi rotation is the pulse-area $\Theta\propto\sqrt{P/\Delta E_c}$, with $P$ the time-averaged incident power. Secondly, the envelope of the oscillation is independent of the spectral width of the control pulse.
Hence, for a given pulse-area, longer time duration pulses suffer less coherence loss than shorter ones. This contradicts a model of coherence decay with rates that are independent of the driving field, where the total loss of coherence should increase with the time duration of the control laser pulse
 \cite{Kyoseva_pra}. Therefore, any model of the observed damping requires an EID mechanism in which the total loss of coherence is a function of the time-integrated power.

To distinguish between phonon and wetting layer models of EID, temperature dependent Rabi rotation measurements were made, and are presented in
Fig.~\ref{fig:RR}(c).
To simplify the ensuing analysis, a spectrally narrow ($\Delta E_{\rm c}=0.2~\mathrm{meV}, ~\tau=4~\mathrm{ps}$) Gaussian pulse is used, where $\tau$ is the time-duration.
The red-shift of the neutral exciton transition with temperature is known to be independent of the dot details \cite{Ortner_prb},
and is used here as an {\it in-situ} thermometer.  At high pulse-areas the photocurrent tends towards a  value
close to half the maximum measured photocurrent. This indicates a final exciton population close to one half, suggesting that the EID
 mechanism does not result in the creation
of additional carriers, since this would lead to a higher photocurrent signal in the high pulse-area limit. Above 30~K, the period of the oscillation increases with temperature by up to 20\%. A possible origin of this variation in the period is discussed briefly later.
As the
temperature increases the damping gets stronger, which is strong evidence for the role played by phonons. The
wetting layer model can be ruled out, since the corresponding rate of EID is proportional to the absorption strength of the wetting layer transitions \cite{VillasBoas_prl}, and any temperature dependence would be characterized by an activation energy of $> 20~\mathrm{meV}$ \cite{Vasanelli_prl}.
The lines show fits to the data, and will be discussed below.

For a quantitative comparison of experiment and theory we consider a coupled dot-phonon model of EID \cite{Forstner_prl,Vagov_prl,Gauger_prb}.
A summary of the model is presented here; further details can be found in ref. \cite{suppl}.
The dot has a crystal ground-state $\vert 0\rangle$ and single exciton state $\vert X\rangle$, addressed
by a resonant control laser of Rabi frequency $\Omega(t)=(\Theta/2\tau\sqrt{\pi})\mathrm{exp}(-(t/2\tau)^{\rm 2})$, where $\tau$ is the pulse width. The pulse spectral width is large compared with the $17~\mathrm{\mu eV}$
neutral exciton fine-structure splitting, and hence the 3-level V-transition of the neutral exciton acts as a two-level transition  on the timescale of the experiment \cite{Boyle_prb}.
The carriers interact with a reservoir of acoustic phonons of wavevector $\mathbf{q}$, for which a linear dispersion, $\omega_{\mathbf{q}}=c_{s}q$, with $c_s$ the speed of sound, is a good approximation in the relevant long wavevector limit. Since piezoelectric coupling is expected to be weak~\cite{Krummheuer_prb}, we focus on deformation potential coupling to LA-phonons as the dominant dephasing mechanism.
Longitudinal optical phonons may also be neglected as their
$\sim 28-36~\mathrm{meV}$ energies are large compared with both
the Rabi energies and temperatures considered here.

In a frame rotating at the laser frequency $\omega_l$, and after a rotating-wave approximation on the driving field, the relevant Hamiltonian may be written $H=H_{C}+H_{B}+H_{I}$. On resonance, the control is $H_{C}=(\hbar\Omega(t)/2)[\vert 0\rangle\langle X\vert +\vert X\rangle\langle 0\vert]$, the phonon bath is described by $H_B=\sum_{\bf q}\hbar\omega_{\bf q}b^{\dagger}_{\bf q}b_{\bf q}$, and the exciton-phonon interaction is written $H_I=\vert X\rangle\langle X\vert \sum_{\mathbf{q}} \hbar( g^*_{\mathbf{q}}b^{\dagger}_{\mathbf{q}}+g_{\mathbf{q}}b_{\mathbf{q}})$,
where $b^{\dagger}_{\mathbf{q}}$ ($b_{\mathbf{q}}$) are creation (annihilation) operators
for bulk acoustic phonons of 
frequency $\omega_{\bf q}$. The deformation potential coupling is given by
$g_{\mathbf{q}}=q(D_{\rm e}{\cal P}[\psi^{\rm e}({\bf r})]-D_{\rm h}{\cal P}[\psi^{\rm h}({\bf r})])/\sqrt{2\mu\hbar\omega_{\mathbf{q}}V}$,
where $D_{\rm e(h)}$ is the deformation potential constant for the electron (hole), $\mu$ is the mass-density of the host material, $V$ the volume of the unit cell, and ${\cal P}[\psi^{\rm e(h)}({\bf r})]$ the form factor of the electron (hole) wavefunction. The model assumes that the lattice properties of the strained dot material (InGaAs) are similar to the host (GaAs) material, so that the wavevector 
is a good quantum number \cite{Krummheuer_prb}. According to the model, the dephasing is dominated by phonons with an energy equal to the Rabi energy. The maximum peak Rabi energy used in these experiments is 0.9 meV, corresponding to LA-phonons of wavelength greater than 23 nm. This is large compared with the 7 to 8-nm FWHM of the carrier wavefunctions of a typical InAs/GaAs dot measured by magnetic tunneling spectroscopy \cite{Patane_prb}, hence ${\cal P}\approx 1$ is a good approximation. 


We describe the phonon-influenced exciton dynamics by a Born-Markov master equation for the system density operator $\rho$, derived ~\cite{Mogilevtsev_prl} under the assumption that the relaxation of the phonon bath, with a half-life of 0.8 to 1.4 ps \cite{Borri_prl}, is fast compared with the exciton dynamics governed by the FWHM of $\Omega(t)$: $\tau_{FWHM}^{(\Omega)}=4\tau\sqrt{\ln{(2)}}=14~\mathrm{ps}$.
We find the exciton coherence $\rho_{0X}=\langle 0 |\rho |X\rangle$ evolves as
\begin{equation}\label{rho0xdynamcis}
\dot{\rho}_{0X}\approx(i\Omega(t)/2)[1-2\rho_{XX}]-[\Gamma_{\rm 2}^*+K_{\rm 2}\Omega^2(t)]\rho_{0X},
\end{equation}
where $\rho_{XX}=\langle X |\rho |X\rangle$ is the population of the excitonic state, obeying
\begin{equation}\label{rhoxxdynamcis}
\dot{\rho}_{XX}=(i\Omega(t)/2)[\rho_{X0}-\rho_{0X}],
\end{equation}
and $\Gamma_{\rm 2}^*$ is a phenomenological, field-independent rate accounting for any additional pure dephasing processes. We see then that the dominant phonon contribution to the dynamics is a driving-dependent dephasing term $\Gamma_2=K(\Omega)\approx K_2\Omega^2(t)$, which arises from the full frequency response of the dot-phonon interaction, $K(\Omega)=[(D_{\rm e}-D_{\rm h})^2\Omega^3/8\pi\mu c_{\rm s}^5\hbar]\coth{(\hbar\Omega/2k_{\rm B}T)}$, in the regime $k_{\rm B}T>\hbar\Omega/2$.
Hence, EID is described by a characteristic timescale $K_{\rm 2}$ that is proportional to temperature, with a gradient $A$ that depends only on bulk material parameters and is independent of dot details:
\begin{equation}
K_{\rm 2}=\frac{(D_{\rm e}-D_{\rm h})^2}{4\pi\mu c_{\rm s}^5\hbar^2}k_{\rm B}T\equiv AT.
\label{eq:K2}
\end{equation}

\begin{figure}
\begin{center}
\vspace{0.2 cm}
\includegraphics[scale=1.0]{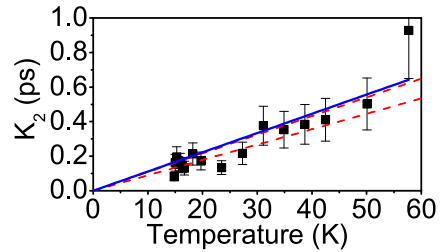}
\vspace{0.2 cm}
\end{center}
\caption{Linear dependence of the excitation-induced-dephasing time $K_{2}$ on temperature: (red-dashed) range of values calculated for bulk GaAs;
(blue-solid) fit to $K_2=AT$.
}\label{fig:D}
\end{figure}

To extract the $K_2$ time from the Rabi rotations in Fig.~\ref{fig:RR}(c),
the data are fitted to a numerical solution of Eqs.~(\ref{rho0xdynamcis}-\ref{rhoxxdynamcis}), using a differential evolution algorithm \cite{website}. This approach to extracting parameters from the data 
follows that described 
by Heiss {\it et~al.} in Ref.~\cite{Heiss_prb}. Excellent fits are obtained to the data in Fig.~\ref{fig:RR}(c), and are presented as red solid lines. The only fitting parameters are $K_{\rm 2}$, $\Gamma_{\rm 2}^*$, and the effective dipole $d=\Theta/\sqrt{P}$.
The effective dipole is a scaling parameter, and exhibits a small, less than 20\%, roll-off with temperature (not shown). This leads to the increase in the period of oscillation with increasing temperature as seen in Fig. \ref{fig:RR}(c). We cautiously suggest that the temperature dependence of the effective optical dipole could be due to Rabi frequency renormalization \cite{Krugel_apb} caused by the dot-phonon interaction. The additional dephasing rate $\Gamma_{\rm 2}^*$ is slow compared with the pulse duration and cannot be accurately determined from the data; the effect of $\Gamma_{2}^*$ is a loss of contrast that is nearly independent of pulse-area~\cite{Kyoseva_pra}.
The fits are most sensitive to $K_{2}$, since this results in a loss of contrast that increases with pulse-area. The dominant source of error is the gradient used for the background subtraction in the raw photocurrent data, as referred to earlier \cite{Stufler_prb}. The error-bars for $K_2$ shown in fig. \ref{fig:D} are estimated by comparing the $K_2$ values extracted from the data for a range of background gradients that still result in good fits to data.

Fig.~\ref{fig:D} presents the temperature dependence of the EID-time
$K_{2}$ extracted from the Rabi rotation data in Fig.~\ref{fig:RR}(c). A near to linear variation is observed. The blue line presents a fit to $K_2=AT$. The value of $A$ extracted from the fit, $A_{\rm meas}=11\pm 1~\mathrm{fs.K^{-1}}$.
Using $A_{\rm meas}$, we estimate an effective band-gap deformation potential of $\vert D_{\rm e}-D_{\rm h}\vert_{\rm QD}=9.0\pm 0.3~\mathrm{eV}$, calculated using Eq. \ref{eq:K2} with bulk GaAs parameters:
$\mu=5.37~\mathrm{g.cm^{-3}}$ and  $c_{\rm s}=5.11~\mathrm{nm.ps^{\rm -1}}$ \cite{Krummheuer_prb,Madelung}. This is close to the value of the hydrostatic deformation potential of the GaAs bandgap reported in the literature, $(D_{\rm e}-D_{\rm h})= -8.5\pm 0.4~\mathrm{eV}$  \cite{Vurgaftman_jap}.
To further show that the data falls within the range expected by the theory, the dashed red-lines show the range of values for $K_2=A_{\rm calc}T$ expected for  $A_{\rm calc}=9.8\pm 0.9~\mathrm{fs.K^{-1}}$, where $A_{\rm calc}$ is calculated using the literature values for the deformation potential given earlier.
The data is in close quantitative agreement with the model, confirming that the dominant source of EID is indeed due to LA-phonons. Further, this suggests that using bulk GaAs phonon modes to calculate the rate of EID is a good approximation, consistent with the calculations  of Ref.~\cite{Grosse_prb}. 


To summarize, we present an experimental study of the intensity damping of
excitonic Rabi rotations of single InGaAs/GaAs dots using photocurrent detection.
We demonstrate strong experimental evidence identifying LA-phonons as the principal source
of intensity damping.
In the high temperature, low Rabi frequency, regime appropriate to our experiments,
the intensity damping can be explained using a dephasing rate with an  excitation induced dephasing term $\Gamma_{2}=AT\Omega^2$.
We measure $A=11 \pm 1 ~\mathrm{fs.K^{\rm -1}}$, in agreement with a first principles calculation of a dot interacting with a reservoir of bulk GaAs LA-phonons. 
The $A$-parameter is a property of the host material only, in this case GaAs.

This work suggests a number of approaches for improving the fidelity of coherent optical control in the solid-state. Most straightforwardly, the rate of EID can be suppressed  by using low temperatures and slower control pulses. Minimizing the exciton population by using off-resonant control schemes should also suppress the dephasing effect~\cite{Gauger_prb}.  It is also notable that diamond has a speed of sound that is 2.35 times faster than that of GaAs, suggesting that NV-centers in diamond may have favorable LA-phonon induced EID-times.

The authors thank the EPSRC (UK) EP/G001642, the QIPIRC UK, and the UK-India Education Research Initiative/ Department of Science
  and Technology (India) for financial support. AN is supported by the EPSRC, and BWL by the Royal Society. We thank H.~Y.~Liu and M.~Hopkinson for sample growth, and D.~M.~Whittaker, and  P.~Kok
  for fruitful discussions.

\bibliographystyle{apsrev}

\end{document}